\documentclass[11pt]{article}
\usepackage{moriond,epsf}
\begin{document}
\vspace*{4cm}
\title{\boldmath First D\O\ Jet Measurements at $\sqrt{s}=1.96$~TeV}

\author{Michael Begel \\
for the D\O\ Collaboration}

\address{Department of Physics and Astronomy, University of Rochester, Rochester NY 14627 USA}

\maketitle\abstracts{
We present the first Run~II measurements with the D\O\ detector of the
inclusive jet and dijet cross sections at $\sqrt{s}=1.96$~TeV.  This
analysis is based on an integrated luminosity of $34\;pb^{-1}$.  The
results from a next-to-leading order perturbative QCD calculation are
compared to the measured cross sections.  The theoretical calculation
agrees with the data.}

\noindent
Measurements of jet production can be used to test QCD predictions for
parton-parton scattering and to constrain the parton density functions
(PDF) of the proton, particularly the gluon distribution~\cite{stump}.
Additionally, they can be used to search for new physics including
quark compositeness, excited quarks, $Z^\prime$s, $W^\prime$s, and
other exotic particles.

D\O\ is a collider detector located at the Fermi National Accelerator
Laboratory near Chicago, Illinois.  The detector has a central
tracking spectrometer consisting of a silicon microstrip tracker and a
scintillating fiber detector located within a $2\,{\rm T}$ solenoidal
magnetic field.  Particle energies are measured in a large
uranium--liquid argon calorimeter supplemented by a scintillator-based
preshower detector.  Muons are detected in several layers of drift
tubes and scintillators sandwiching a toroid magnet.  Roman pot
detectors have been deployed within the accelerator lattice to measure
scattered protons and anti-protons.  This is the second run for the
D\O\ detector which has significant upgrades to the tracking systems
and electronics compared with the previous run.  The current
center-of-mass energy of the Tevatron accelerator is 1.96~TeV.  This
is $\approx9$\% larger than the $\sqrt{s}$ for the previous run, but
leads to a significant increase in the jet cross section at high-$p_T$
(a factor of~2 increase at $p_T\approx400$~GeV).

Outgoing partons from the hard scattering process hadronize to form
jets of particles.  These jets are measured in the D\O\ calorimeter.
Jets are reconstructed using an iterative jet cone algorithm with
$R=0.7$ ($R^2 = \Delta\eta^2 + \Delta\phi^2$).  The calorimeter energy
is corrected back to the particle level using information from $\gamma
+ {\rm jet}$ events, low bias triggers, and Monte Carlo simulations.
There are large statistical uncertainties and substantial systematic
uncertainties in this energy scale determination that increase with
energy due to extrapolation.  These are principally caused by small
$\gamma + {\rm jet}$ statistics above 200~GeV.  This is the dominant
systematic uncertainty in the jet cross section measurements.  We
restricted the jet measurements to $|\eta|<0.5$ to limit the impact of
these uncertainties.

\begin{figure}
\vskip-0.25truein
\epsfxsize=3truein
\begin{center}
\epsffile{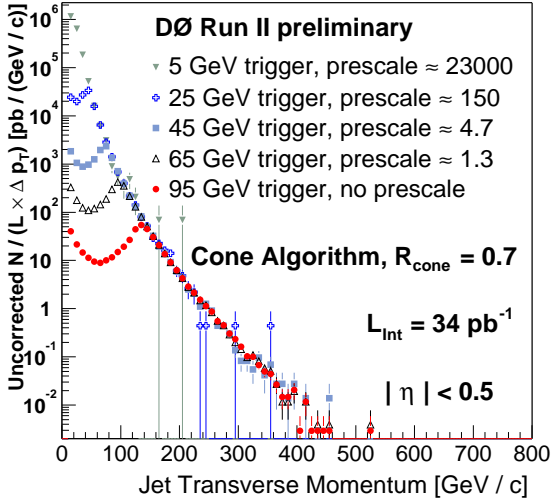}
\end{center}
\caption{The uncorrected inclusive jet normalized $p_T$ distribution for 
five jet triggers.
\label{fig:inctrig}}
\end{figure}

Events chosen for this analysis required a good primary vertex
($\epsilon\approx78$\%) and high quality jets where the jet selection
was based on calorimeter characteristics to reduce fakes and noise
($\epsilon\approx97$\% per jet).  The total integrated luminosity
considered in this analysis was $34\;pb^{-1}$.  Four inclusive jet
triggers were used in this study.  Each trigger required a localized
energy deposited in the calorimeter and a ``reconstructed'' jet with
$p_T$ above a threshold: 25, 45, 65, or 95~GeV.  These thresholds, and
the corresponding prescales, were chosen to span a wide $p_T$ range.
The uncorrected jet $p_T$ spectrum, normalized to the exposed
luminosity (accounting for the prescales), is shown in
Fig.~\ref{fig:inctrig}.  The figure also includes the jet spectrum
from a trigger that only required a localized deposition of energy in
the calorimeter.  This trigger was used to study the lowest threshold
trigger.  These inclusive triggers were also used in the studies of
dijet production.

\begin{figure}
\vskip-0.25truein
\begin{center}
\hbox{\vtop{\hsize=3truein
\epsfxsize=3truein
\epsffile{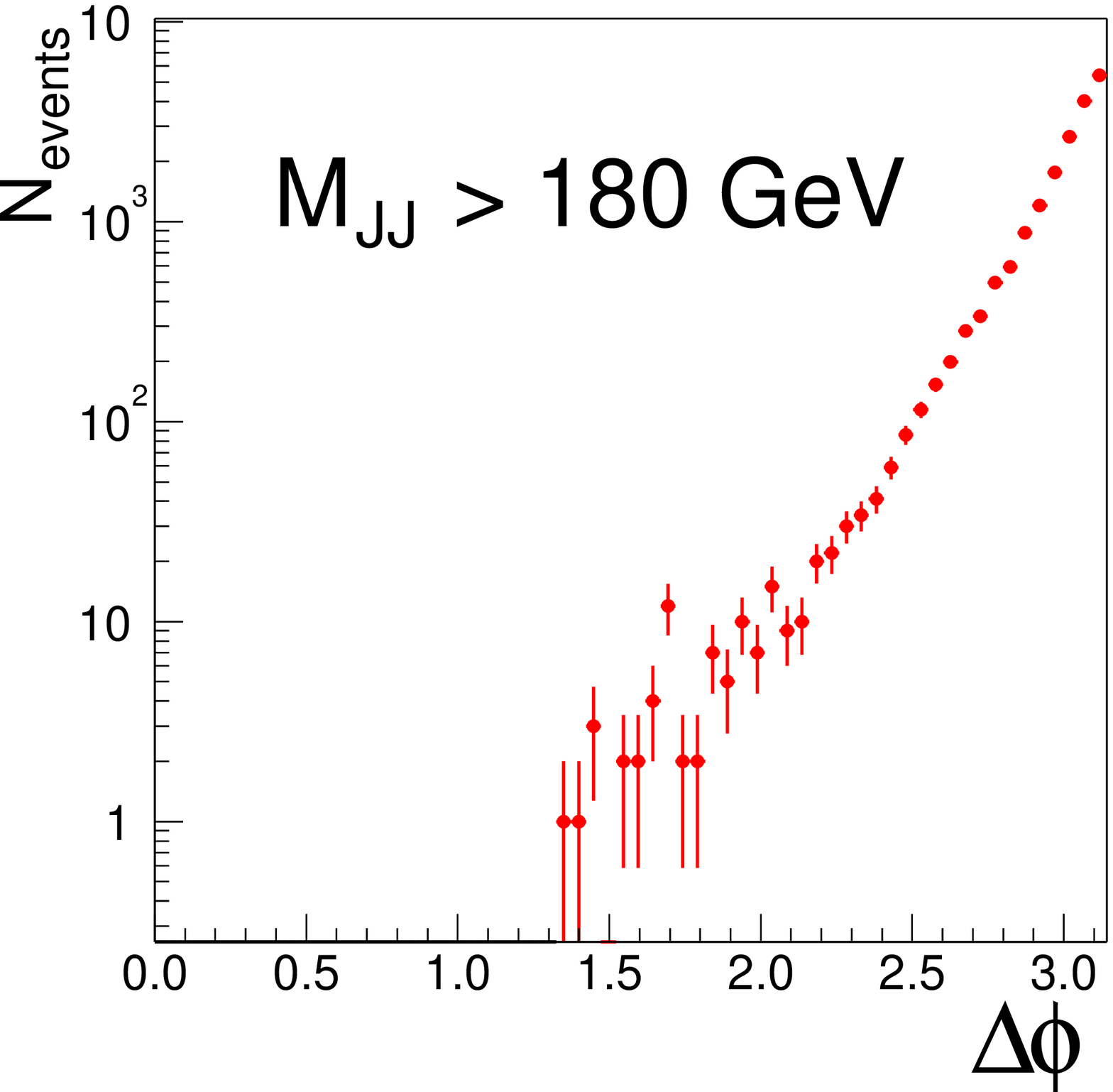}}
\vtop{\hsize=3truein
\epsfxsize=3truein
\epsffile{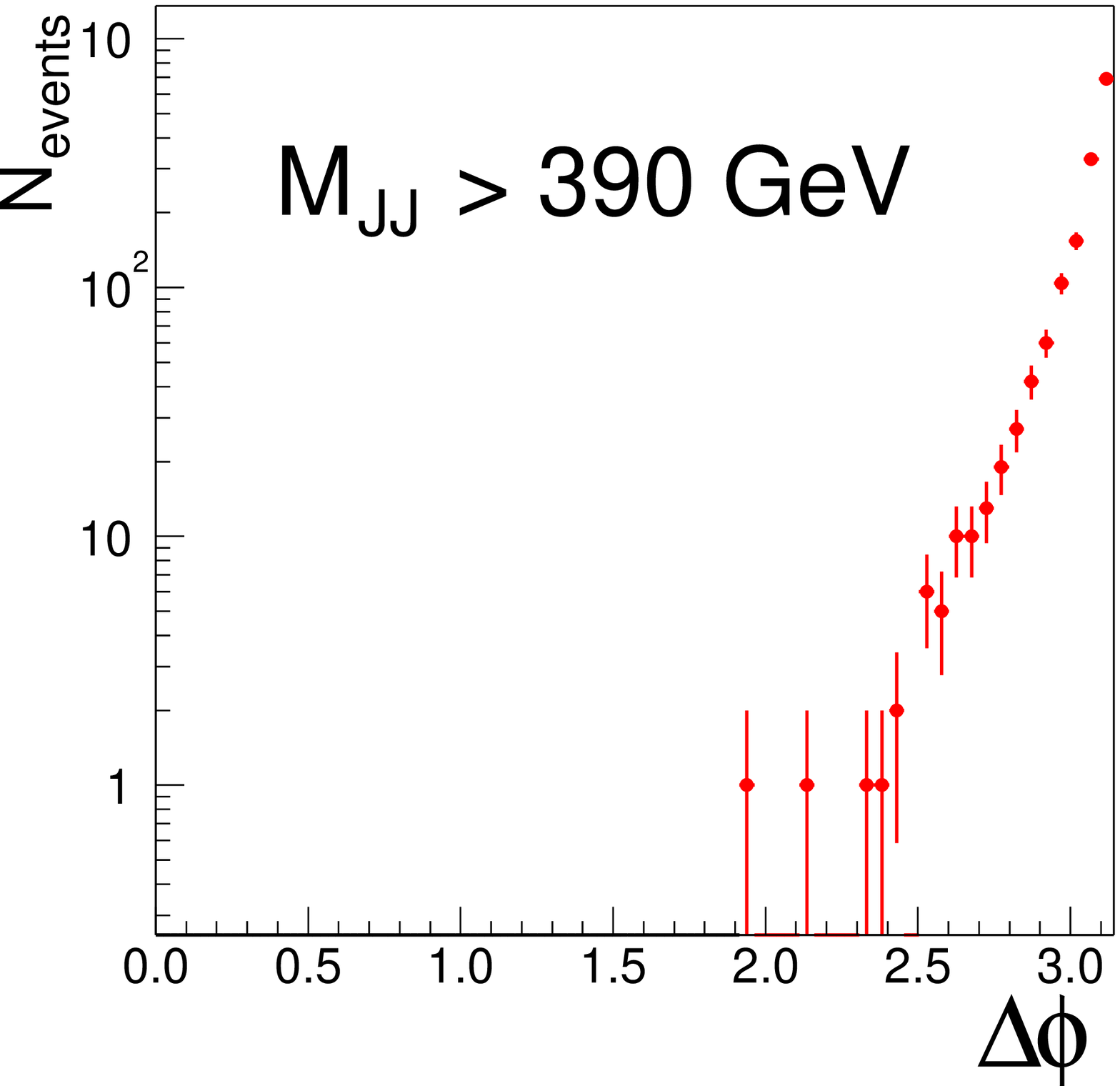}}}
\end{center}
\caption{The $\Delta \phi$ distribution for dijets with 
$M_{JJ}>180$~GeV (left) and $M_{JJ}>390$~GeV (right).
\label{fig:deltaphi}}
\epsfxsize=3truein
\begin{center}
\hbox{\vtop{\hsize=3truein
\epsffile{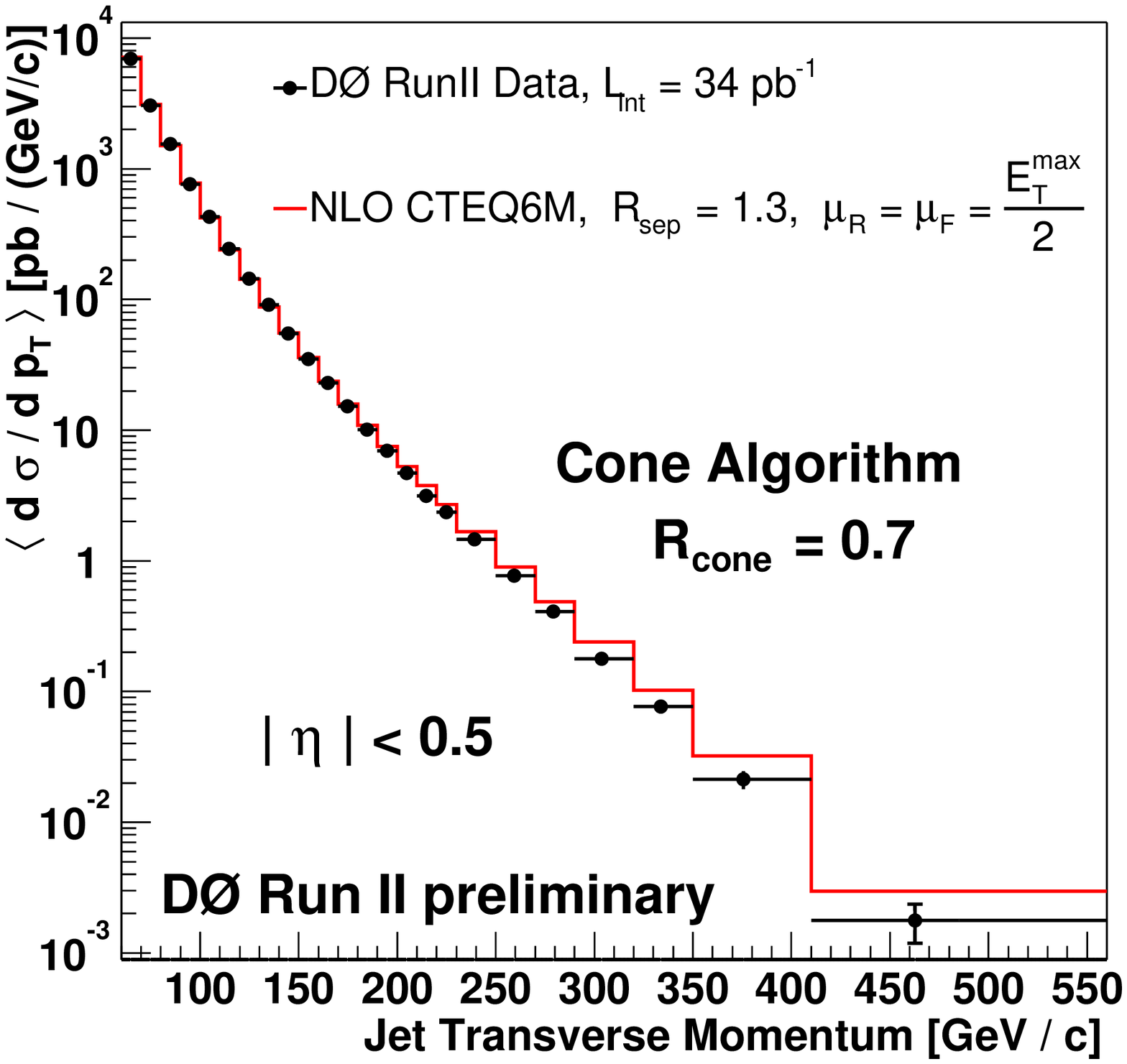}}
\vtop{\hsize=3truein
\epsfxsize=3truein
\epsffile{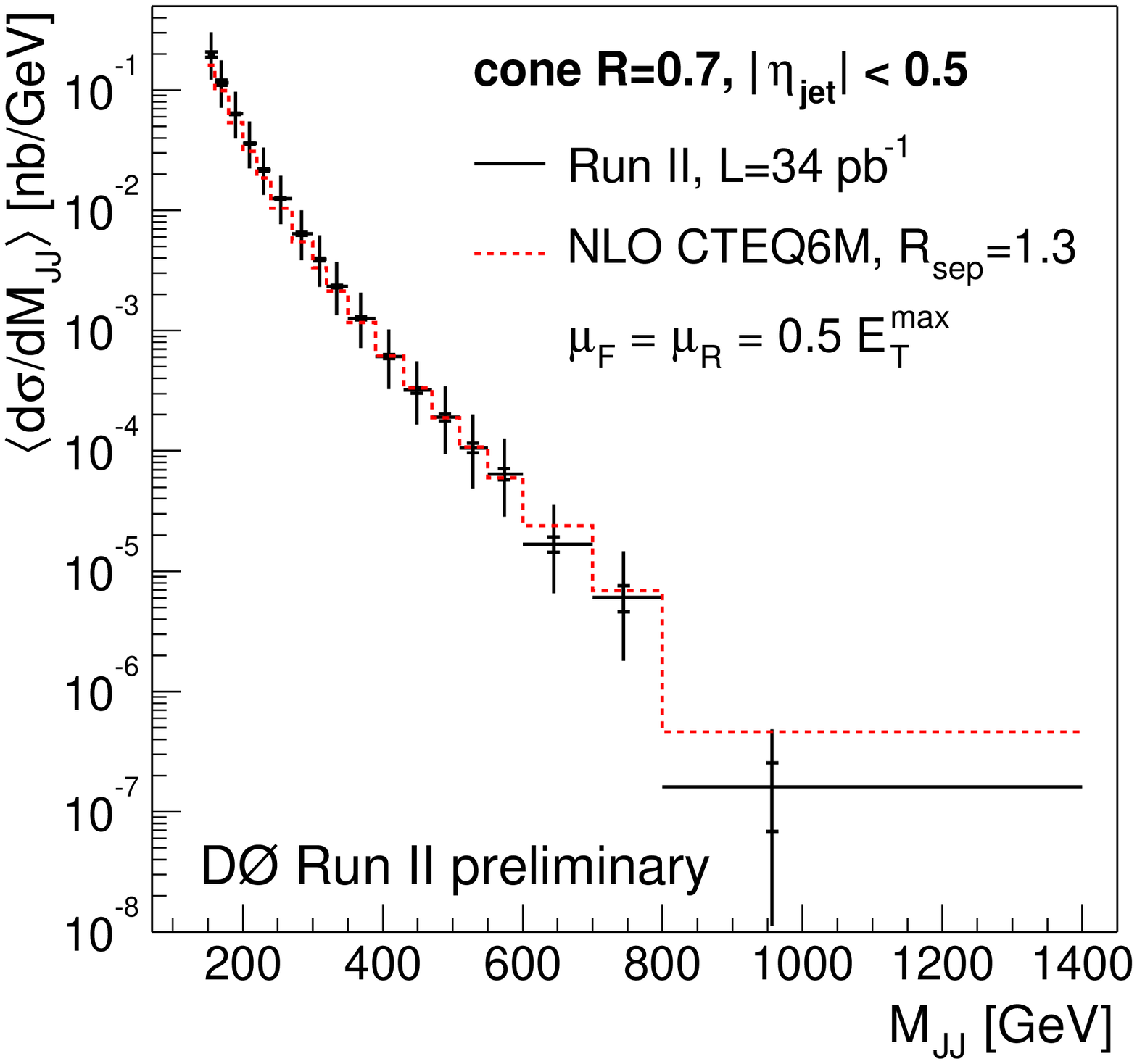}}}
\end{center}
\caption{Left: the inclusive jet cross section as a function of $p_T$.  
Right: the dijet cross section as a function of dijet mass.  Overlayed
on the data are the predictions of a NLO pQCD calculation.
\label{fig:logxs}}
\end{figure}

The dijet sample was defined by events with at least two jets where
the two leading jets with the highest $p_T$ were required to have
$|\eta|<0.5$.  The difference in the azimuthal angle of the two
leading jets for two inclusive mass ranges is shown in
Fig.~\ref{fig:deltaphi}.  As expected, most jets in this sample are
balanced in $\phi$.  The jet $p_T$ resolution was measured from the
$p_T$ imbalance in these events.  This resolution was used to estimate
the unsmearing correction applied to both the inclusive jet $p_T$
spectrum ($\approx10$--15\%) and the dijet mass spectrum
($\approx5$--15\%).

The inclusive jet differential cross section as a function of $p_T$ is
shown in Fig.~\ref{fig:logxs}~left.  The dijet differential cross
section as a function of mass is shown in Fig.~\ref{fig:logxs}~right.
Overlayed on the data in Fig.~\ref{fig:logxs} are the results of a NLO
pQCD calculation {\tt JETRAD}~\cite{jetrad} using the CTEQ6M
PDF~\cite{cteq6}.  The factorization and renormalization scales have
been set equal to $p_T/2$ of the leading jet in the event and
$R_{sep}$ has been set to 1.3.  Reasonable alterations to the scale
choice or $R_{sep}$ parameter lead to $\approx10$\% changes in the
calculation.

\begin{figure}
\vskip-0.25truein
\begin{center}
\hbox{\vtop{\hsize=3truein
\epsfxsize=3truein
\epsffile{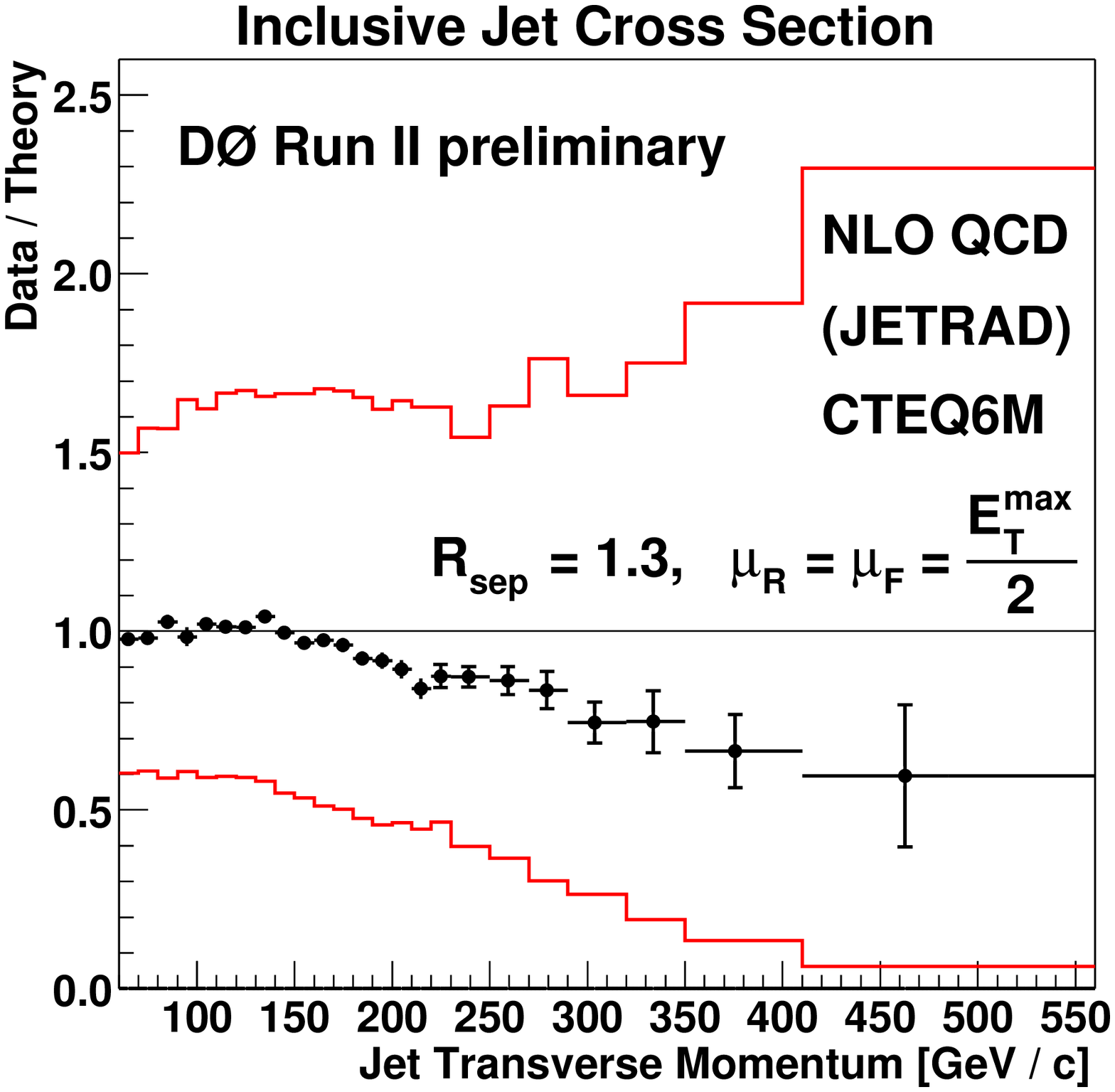}}
\vtop{\hsize=3truein
\epsfxsize=3truein
\epsffile{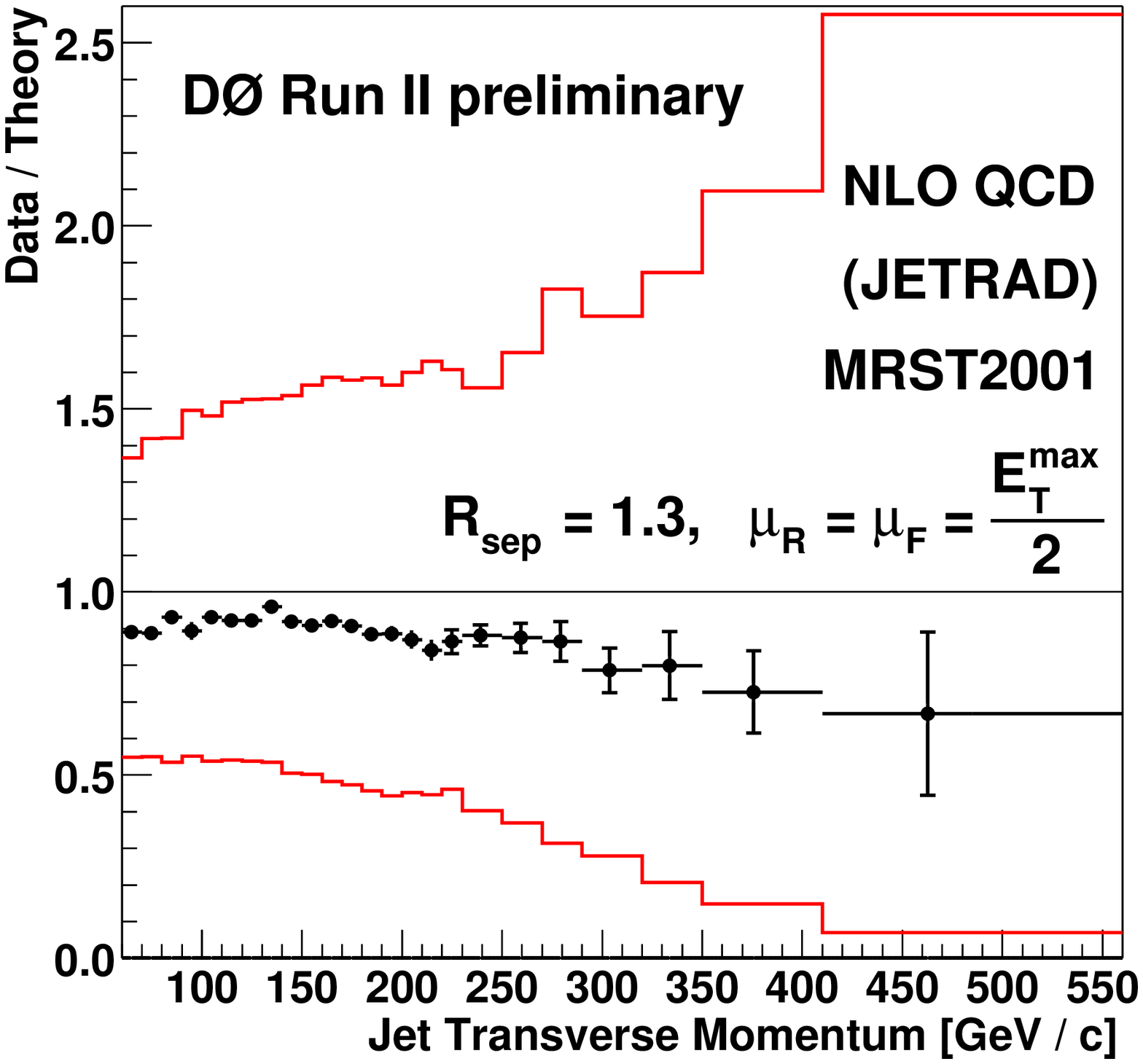}}}
\end{center}
\caption{The inclusive jet cross section shown as data/theory as a function 
of $p_T$.  Two PDFs were used in the theory calculation: CTEQ6M~(left)
and MRST2001~(right).
\label{fig:incxslin}}
\begin{center}
\hbox{\vtop{\hsize=3truein
\epsfxsize=3truein
\epsffile{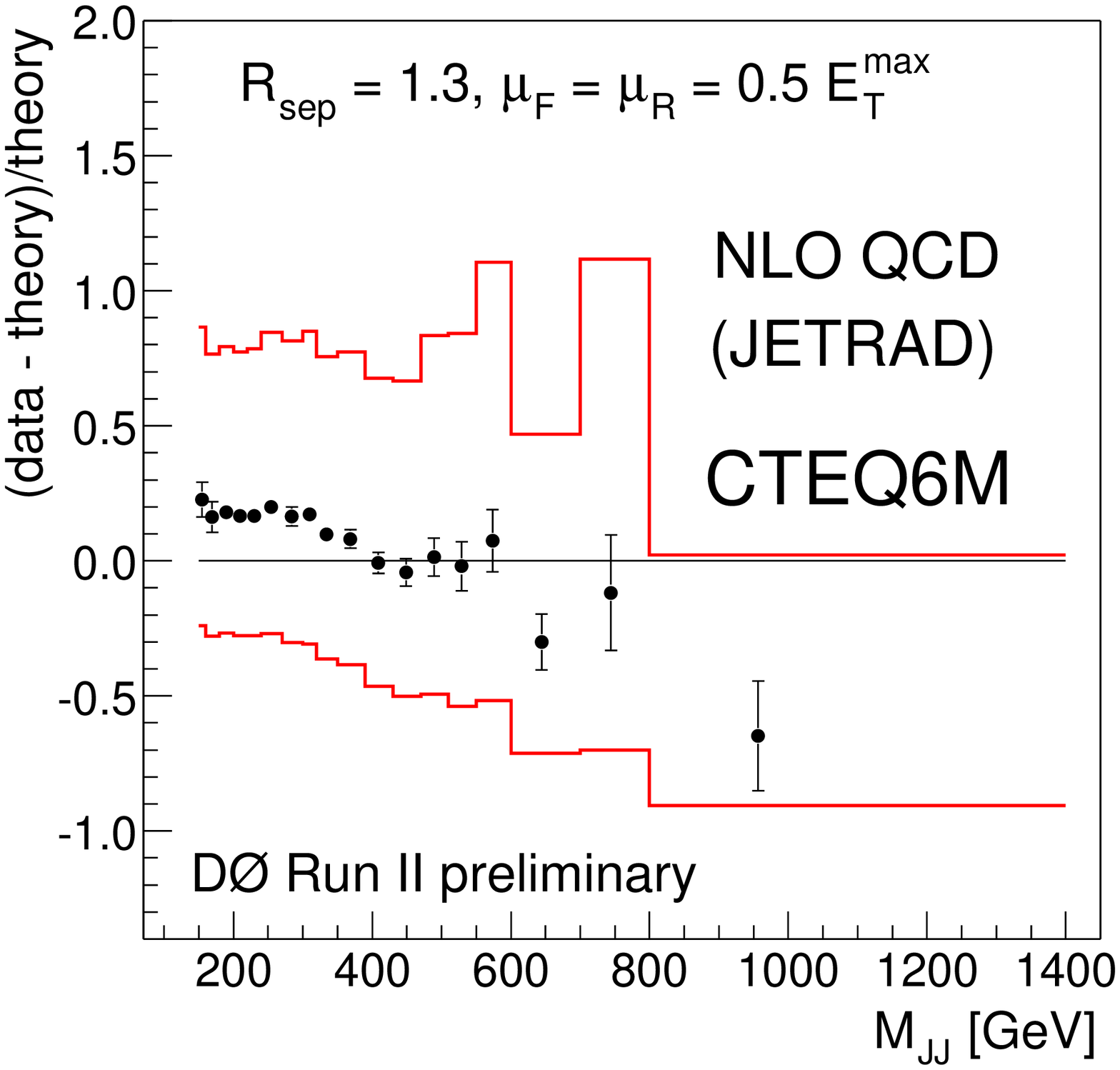}}
\vtop{\hsize=3truein
\epsfxsize=3truein
\epsffile{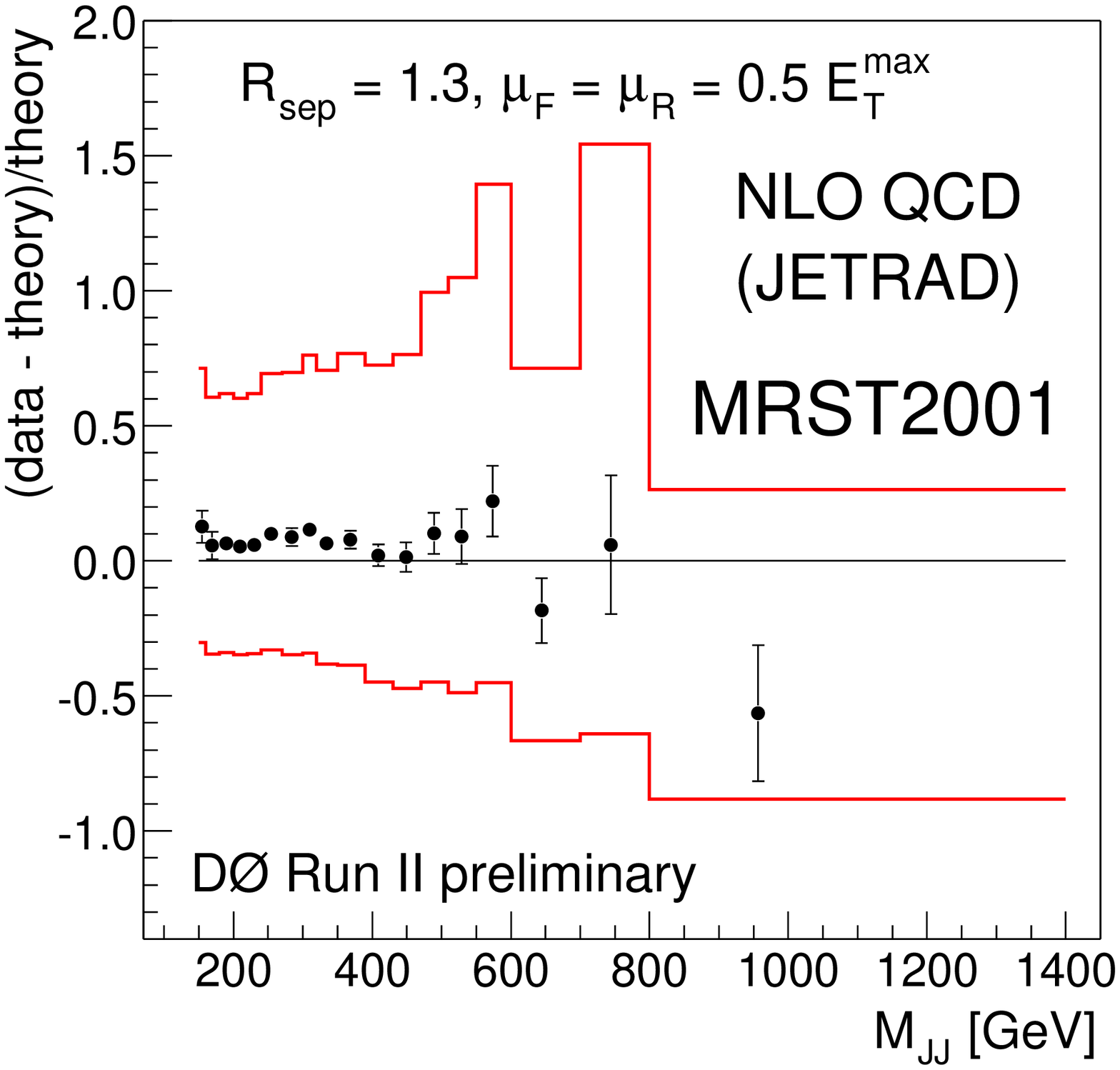}}}
\end{center}
\caption{The dijet cross section shown as (data -- theory)/theory as a function 
of $M_{JJ}$.  Two PDFs were used in the theory calculation: CTEQ6M~(left)
and MRST2001~(right).
\label{fig:dijetxslin}}
\end{figure}

Linear comparisons of the calculation to the data are presented in
Figs.~\ref{fig:incxslin} and~\ref{fig:dijetxslin}.  Here we show {\tt
JETRAD} using both CTEQ6M and MRST2001 PDF~\cite{mrst}.  The theory is
well within the data uncertainties (dominated by the energy scale
calibration).  While the uncertainties are too large to prefer a PDF,
it is interesting to note the different $p_T$ dependence of the
calculations due to PDF selection.

We presented the first Run~II measurements of the inclusive jet and
dijet cross sections with the D\O\ detector at $\sqrt{s}=1.96$~TeV.
This analysis was based on an integrated luminosity of $34\;pb^{-1}$.
We have accumulated much more luminosity than was used in this study
and plan to update these results with increased $\eta$ coverage and
decreased energy scale uncertainties in the near future.

\end{document}